\documentclass{webofc}
\usepackage[varg]{txfonts}   
\begin{document}
\title{Observing galaxy clusters and the cosmic web through the Sunyaev Zel'dovich effect with MISTRAL}
%
%

\author{\lastname{E.S. Battistelli}\inst{1,2,3}\fnsep\thanks{\email{elia.battistelli@roma1.infn.it}} 
\and \lastname{E. Barbavara}\inst{1} \and \lastname{P. de Bernardis}\inst{1} \and \lastname{F. Cacciotti}\inst{1} \and
        \lastname{V. Capalbo}\inst{1} \and \lastname{A. Carbone}\inst{1} \and \lastname{E. Carretti}\inst{4} \and
        \lastname{D. Ciccalotti}\inst{1} \and \lastname{F. Columbro}\inst{1} \and \lastname{A. Coppolecchia}\inst{1} \and
        \lastname{A. Cruciani}\inst{3} \and \lastname{G. D'Alessandro}\inst{1} \and \lastname{M. De Petris}\inst{1} \and
         \lastname{F. Govoni}\inst{5} \and
        \lastname{G. Isopi}\inst{1} \and \lastname{L. Lamagna}\inst{1} \and \lastname{E. Levati}\inst{1} \and
         \lastname{P. Marongiu}\inst{5} \and \lastname{A. Mascia}\inst{1} \and
         \lastname{S. Masi}\inst{1} \and \lastname{E. Molinari}\inst{5} \and \lastname{M. Murgia}\inst{5}  \and
         \lastname{A. Navarrini}\inst{5} \and  \lastname{A. Novelli}\inst{1} \and
         \lastname{A. Occhiuzzi}\inst{1} \and \lastname{A. Orlati}\inst{4} \and
        \lastname{E. Pappalardo}\inst{1} \and \lastname{A. Paiella}\inst{1} \and
        \lastname{G. Pettinari}\inst{6} \and \lastname{F. Piacentini}\inst{1} \and \lastname{T. Pisanu}\inst{5} \and
        \lastname{S. Poppi}\inst{5} \and
        \lastname{I. Porceddu}\inst{5} \and  \lastname{A. Ritacco}\inst{5} \and \lastname{M.R. Schirru}\inst{5}  \and
         \lastname{G.P. Vargiu}\inst{5} 
}

\institute{Sapienza University of Rome, Physics Department, Piazzale Aldo Moro 5,
I-00185 Rome, Italy  
\and
           INAF-IAPS, Via Fosso de Cavaliere, 100. I-00133, Rome, Italy 
\and
           INFN-Sezione di Roma, Piazzale Aldo Moro 5, I-00185, Rome, Italy
\and
           INAF-Istituto di Radioastronomia - Via P. Gobetti, 101, I-40129 Bologna, Italy
\and
           INAF-Osservatorio Astronomico di Cagliari, Via della Scienza 5, I-09047 Selargius (CA), Italy
\and
           Istituto di Fotonica e Nanotecnologie - CNR, Via del Fosso del Cavaliere 100, I-00133 Rome, Italy
          }

\abstract{%
Galaxy clusters and surrounding medium, can be studied using X-ray bremsstrahlung emission and Sunyaev Zel'dovich (SZ) effect. Both astrophysical probes, sample the same environment with different parameters dependance. The SZ effect is relatively more sensitive in low density environments and thus is useful to study the filamentary structures of the cosmic web. In addition, observations of the matter distribution require high angular resolution in order to be able to map the matter distribution within and around galaxy clusters. MISTRAL is a camera working at 90GHz which, once coupled to the Sardinia Radio Telescope, can reach $12''$ angular resolution over $4'$ field of view (f.o.v.). The forecasted sensitivity is $NEFD \simeq 10-15mJy \sqrt{s}$ and the mapping speed is $MS= 380'^{2}/mJy^{2}/h$. MISTRAL was recently installed at the focus of the SRT and soon will take its first photons.}
\maketitle
\section{Introduction}
\label{intro}

The Cosmic Microwave Background (CMB) represents one the most unique source of cosmological information. Studying the primary anisotropies and the polarization of the CMB is allowing us to enter into the so called precision cosmology. Within this framework, we can derive the cosmological parameters with extreme precision and know the energy content of our universe to a fraction of a percent \cite{planck2018a,battistelli2022}.

On the other hand, the nature and the physics of most of the energy content of our universe are still unknown. 68.3\% of the energy content of our universe is in the form of dark energy which is responsible for the acceleration of the universe. 26.8\% of it is in the form of dark matter which can only interact gravitationally with the remaining baryonic matter. In addition, the observed baryonic matter in the local universe is still small compared to what is predicted by nucleosynthesis and by measurements of the CMB power spectrum (see, e.g.,  \cite{planck2018a}). A diffuse baryonic dark matter (missing baryons) could explain, at least in part, the apparent discrepancy between observations and cosmological estimation \cite{martin2023}.

Hydrodynamical simulations of large-scale structures (see, e.g., \cite{cen06}) show that at low redshifts, these missing baryons should lie in the temperature range of 10$^{5}$<T<10$^{7}$ K in a state of warm-hot gas not yet observed through their soft X-ray emission. This warm-hot intergalactic medium (WHIM) is arranged in the form of filamentary structures of low-density intergalactic medium connecting (and around) the clusters of galaxies into the so called cosmic web. 

\section{The Sunyaev Zel'dovich effect in galaxy clusters and in filaments}
\label{szgeneral}

\subsection{Thermal Sunyaev Zel'dovich effect}
\label{sz}

It is well known that the CMB has an almost perfect black body spectrum. However, when the CMB photons scatter off hot electrons present in the Inter Cluster Medium (ICM) present in galaxy clusters, they undergo inverse Compton scattering resulting in a distortion of its frequency spectrum.

This effect (the Sunyaev Zel'dovich, SZ, effect \cite{SZ72}) is due to the energy injection originated by the hot electron gas in galaxy clusters and the surrounding medium. This secondary anisotropy effect produces a brightness change in the CMB that can be detected at millimeter and submillimeter wavelengths, appearing as a negative signal (with respect to the average CMB temperature) at frequencies below $\simeq$217GHz and as a positive signal at higher frequencies. The SZ intensity change directly depends on the electron density of the scattering medium, $n_{e}$, and on the electron temperature $T_{e}$, both integrated over the line of sight $l$, and its spectrum can be described by the following differential intensity:

\begin{equation}
\frac{\Delta I(x)}{I_0} = y \frac{x^4e^x}{(e^x-1)^2}\left(x\coth \frac{x}{2}-4 \right)= yg(x) 
\end{equation}
where: $I_0 = \frac{2h}{c^2}\left(\frac{k_b T_{CMB}}{h} \right)^3 $, $T_{CMB}$ is the CMB temperature, $x=\frac{h\nu}{k_{b}T_{CMB}}$ is the adimensional frequency, and $y=\int{n_{e}\sigma_{T}\frac{k_{B}T_{e}}{m_{e}c^{2}}dl}$ is the Comptonization parameter, $\sigma_{T}$ is the Thomson cross section, $k_{B}$ is the Boltzman constant, $m_{e}$ is the electron mass, and $c$  is the speed of light in vacuum.  The Comptonization parameter $y$ is the integral along the line of sight $l$ of the electron density $n_{e}$ weighted by the electron temperature $T_{e}$ and is the quantity that quantifies the SZ effect: it can be seen as the integrated pressure over the galaxy clusters.

It turns out that the same electrons that scatter off the CMB photons in galaxy clusters, also emit in the X-ray by bremsstrahlung. The bremsstrahlung emission depends on $n_{e}$ and on $T_{e}$ with different dependencies with respect to the SZ effect. In particular, X-ray emission is proportional to $n_{e}^{2}$ and thus, the SZ effect, which is proportional to $n_{e}$, is more sensitive to low density regions. For this reason, it was proposed to use the SZ for low density environments such as the outskirts of galaxy clusters and the filamentary structures between them. 

\subsection{Matter distribution}
\label{distr}

Matter distribution in our universe is clearly non uniform and hydrodynamical simulations predict that matter is distributed in a so-called cosmic web distribution. Simulations can test how structures form and thus investigate the interplay between baryonic matter, dark matter and dark energy. Focussing on a few $Mpc$ scale, allows us to track the progenitor of a group of galaxies or galaxy clusters. Small-mass objects form first at z>5, and quickly grow in size and violently merge with each other, creating increasingly larger and larger system. Hydrodinamical simulation of pre-merging pair adapted to Comptonization parameter $y$ observable, show observable over-densities at angular resolution ranging from $arcmin$ to tens' of $arcsec$ \cite{vazza18}.
This drives to the necessity to observe SZ with high angular resolution, without loosing large scales, and with high sensitivity ($10''$ resolution with few $arcmin$ f.o.v.).

\section{MISTRAL receiver}
\label{mis}

The MIllimeter Sardinia radio Telescope Receiver based on Array of Lumped elements kids (MISTRAL), is a cryogenic camera working at 90GHz between 74GHz and 103GHz. It takes radiation from the $64m$ Sardinia Radio Telescope. MISTRAL hosts an array of 415 Kinetic Inductance Detectors (KIDs) and will measure the sky with $12''$ angular resolution over $4'$ f.o.v.. MISTRAL has recently started its commissioning phase and in 2024, it will start its operations as part of the renewed SRT receiver fleet, as facility instrument.


The Sardinia Radio Telescope (SRT)  \cite{bolli2015}, is a Gregorian configured, fully steerable, 64m-primary mirror radio-telescope which can work from 300MHz to 116GHz. It is a multipurpose instrument with a wide variety of applications which started its scientific programs in 2016. In 2018, a National Operational Program (PON) grant was assigned to INAF with the aim to exploit to the full the SRT capability to reach mm wavelenghts up to 116GHz\cite{Govoni2021}. Among other scientific upgrades, one of the working packages includes a millimetric camera working, in continuum, at 90GHz$\pm$15GHz: MISTRAL receiver, which was built at Sapienza University\cite{batt2023,dale22}. 



\subsection{MISTRAL cryogenic system}
\label{crio}

MISTRAL is a cryogenic camera hosting refocussing optics and an array of Kinetic Inductance Detectors (KIDs). Our KIDs are superconducting detectors made out of Titanium-Aluminium (Ti-Al) bilayer. The critical temperature $T_{c}$ of this alloy is 945mK and thus the detectors have to be cooled down to temperatures well below $T_{c}$. This, in addition to the necessity to cool down the detectors to reduce noise, makes MISTRAL a fairly complicated cryogenic camera. MISTRAL employs a Sumitomo 1.5W Pulse Tube cryocooler\footnote{https://www.shicryogenics.com} and a twin Helium 10 close cycle refrigerator\footnote{https://www.chasecryogenics.com/} and was assembled in UK by QMC instruments\footnote{http://www.qmcinstruments.co.uk/}.

One of the biggest challenges of MISTRAL, is the necessity to work in the Gregorian room of the SRT.  This implies that the receiver will move with the telescope and thus the cryostat will not be steady nor in the vertical position as usually cryogenic equipment need to stay. This has two consequences: a) the insertion of the Pulse Tube head and the refrigerator into the cryostat, is such that they both work in the nominal vertical position when the telescope points at an elevation of 57.5$^{\circ}$. b) The compressor which ensures the operation of the cryocooler has to be put in a position which does not change its inclination. This is possible only in the compressor room which is 120m apart from the Gregorian room. The possibility to have the cryocooler working at such a distance with 120m flexible helium lines was previously tested and proved to be feasible although with some loss of efficiency \cite{coppo23}. In such a way, MISTRAL has been tested to work properly in the inclination range +/-25$^{\circ}$, resulting in elevation range of 32.5-82.5$^{\circ}$ with no degradation of the thermal performance.

\begin{figure}[h]
    \centering
    \includegraphics[width=105mm]{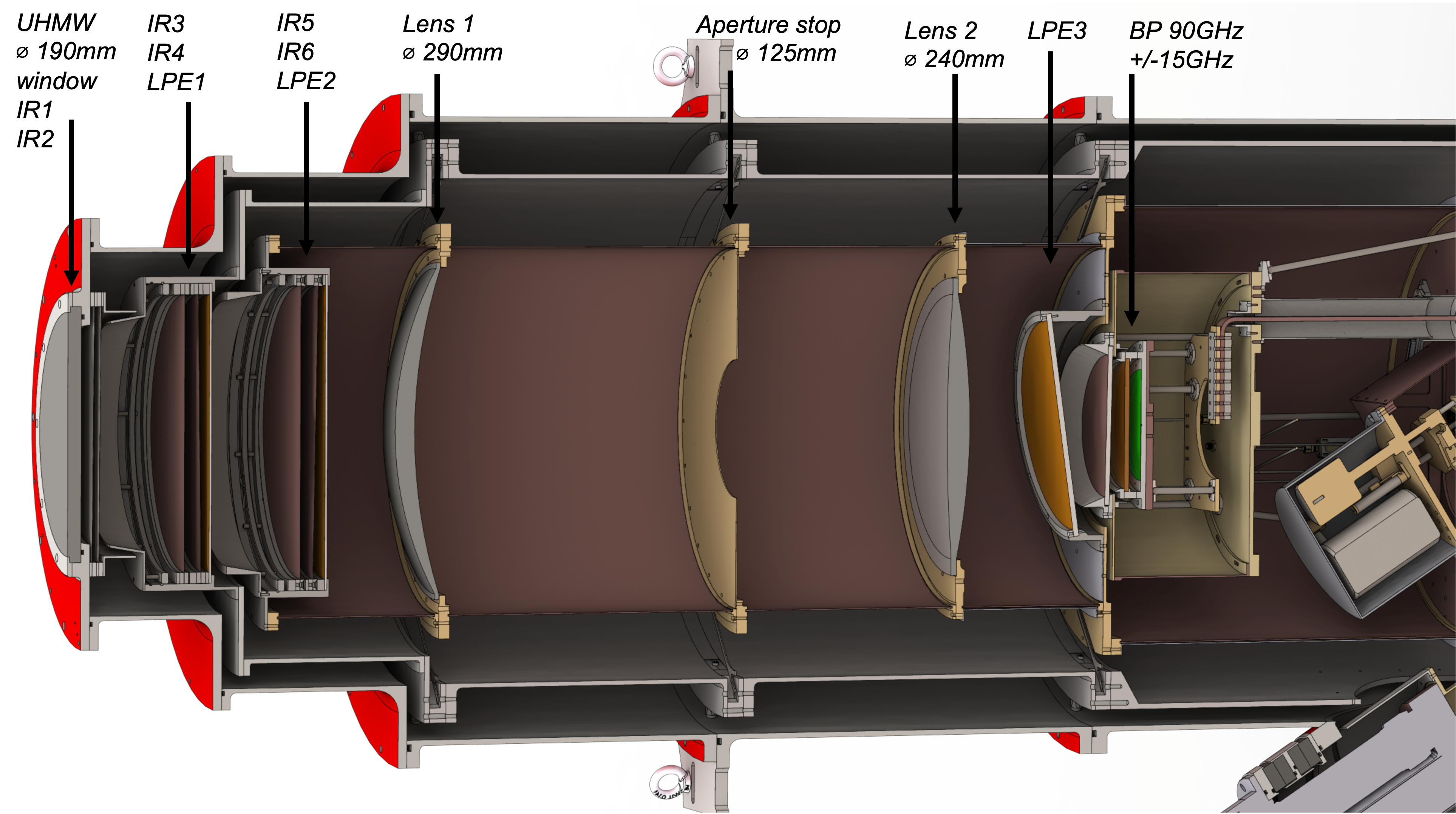}
    \caption{A cut of MISTRAL cryostat highlighting the optics of the receiver: a Ultra High Molecular Weight (UHMW) polyethylene window \cite{dale18} is followed by two infrared (IR) filters. The band selection is actuated by a sequence of Low Pass Edge filters (LPE) and a final Band Pass (BP) filter.}
    \label{fig:cryo}
\end{figure}

\subsection{MISTRAL optics}
\label{opt}

The optical design of MISTRAL includes two Anti Reflection Coated (ARC) Silicon lenses able to image the Gregorian Focus on the array of detectors. Detectors are coupled to radiation through open space (filled array) so, a cryogenic cold stop placed in between the two lenses, is needed to reduce background and avoid stray-light. The bandwidth of operations, as well as the reduction of the load onto the different stages of the cryostat, is provided by a set of radiation quasi-optical filters produced by QMC instruments\footnote{\url{http://www.terahertz.co.uk/qmc-instruments-ltd}}, anchored at the different thermal stages of the cryostat (see Fig. \ref{fig:cryo}).

The two Silicon lenses allow to report $4'$ of the SRT focus onto the array of 415 KIDs. They are anti-reflection coated with Rogers RO3003\footnote{https://www.rogerscorp.com/}. Their diameter is 290mm and 240mm respectively while the aperture cold stop diameter of 125mm. All the lenses+cold stop system is kept at 4K. The in-band average simulations report excellent values with a Strehl Ratio from 0.97 to 0.91 for on-axis and edge positions. Analogously, the FWHM is $12.2''$ on axis, and $12.7''$ at 45mm off axis (which corresponds to 2' in the sky).

\subsection{MISTRAL detectors}
\label{kids}

\begin{figure}[h]
    \centering
    \includegraphics[width=130mm]{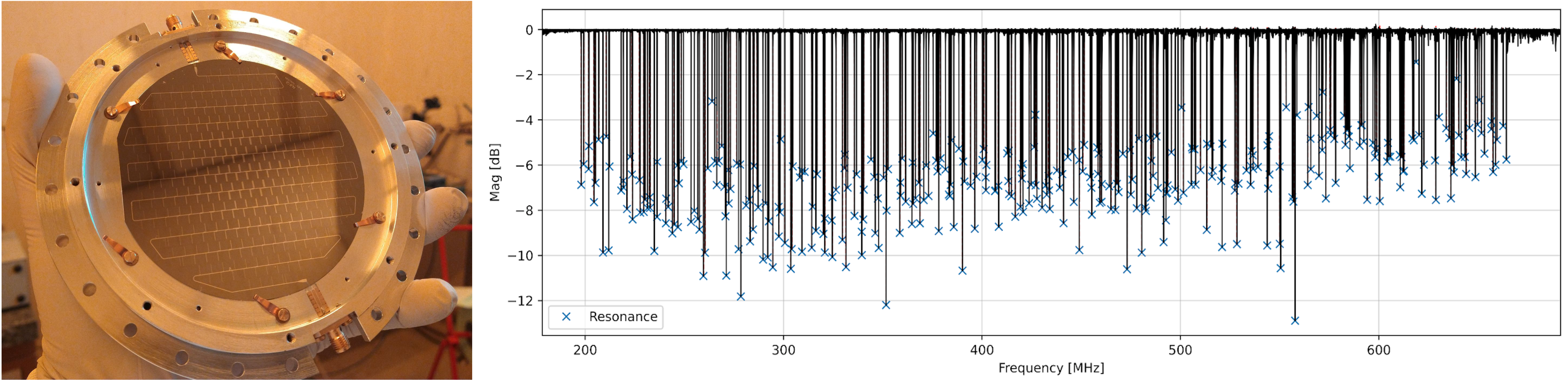}
    \caption{Left: MISTRAL array of KIDs in its holder\cite{Paiella2022}. Right: An image of the response of the KIDs to the tones generated by the ROACH2 electronics sent to MISTRAL.}
    \label{fig:KIDs}
\end{figure}

MISTRAL takes advantage of the high sensitivity of Kinetic Inductance Detectors (KIDs) as well as the capability to Frequency Domain Multiplexing such resonators\cite{Paiella2022,Paiella2023,cacciotti2023}. MISTRAL KIDs are Ti-Al bilayer of thickness 10 + 30 $nm$ with critical temperature $T_{c}=945mK$ and are fabricated at CNR-IFN\footnote{https://ifn.cnr.it/where-we-are/roma/} on 4'' silicon wafer \cite{Paiella2016,Coppolecchia2020} (see Fig. \ref{fig:KIDs}). The feedline is made of Aluminium of 21nm with a critical temperature $T_{c}=1.1K$. This was done to reduce its sensitivity to millimetric radiation. The 415-detectors array is arranged in such a way the each KID  samples the focal plane with angular spacing of $10.6''$, lower than the each pixel angular resolution, thus oversampling the f.o.v.. We use ROACH2 based electronics\footnote{https://casper.astro.berkeley.edu/wiki/ROACH2} to manage the Frequency Domain Multiplexing readout, and to send the tones to bias each of the resonators. 

\section{MISTRAL calibration, installation, and sensitivity forecast}
\label{cal}

MISTRAL has undergone diffuse laboratory calibration, noise measurements, pixel recognition, which certified the health of the instrument. The electric characterization has started with the tuning of the KIDs, the choice of the resonant frequencies and the adjustment of the power to be sent to each KID. Our KIDs are designed to work between 200MHz and 800MHz. The resulting tones are spaced with an averaged separation of 0.96MHz (see Fig. \ref{fig:KIDs}, right panel).  

\begin{figure}[h]
    \centering
    \includegraphics[width=120mm]{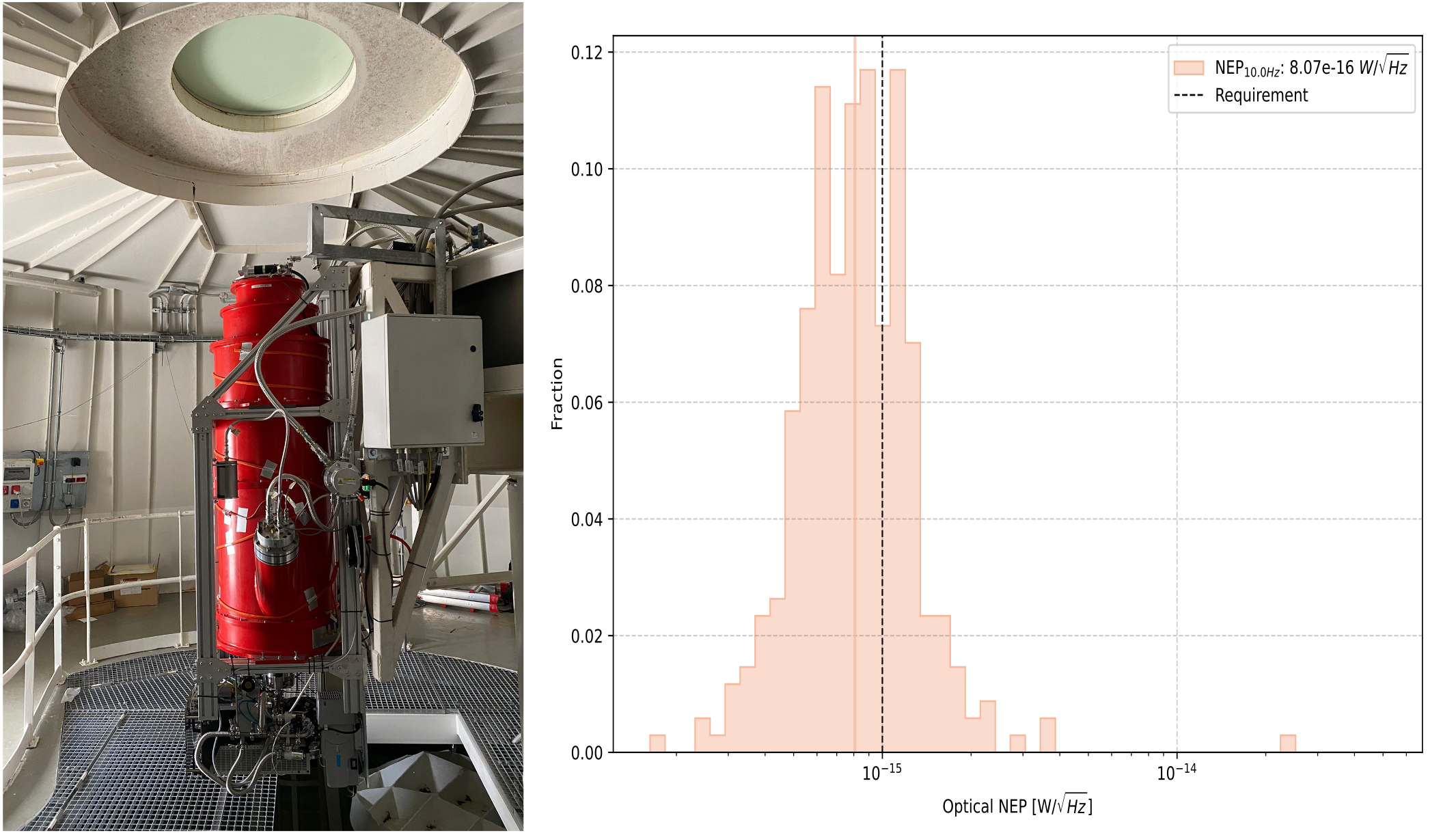}
    \caption{Left: MISTRAL receiver installed in the Gregorian room of the SRT. Right: an histogram of the optical NEP of MISTRAL KIDs. The dashed line represents the limit we were given comparing the optical NEP with the photon noise at the SRT.}
    \label{fig:cal}
\end{figure}

The optical performance have then been measured using an artificial source and a custom designed optical system which sends to MISTRAL KIDs, millimetric radiation with the same beam divergence (i.e. same f/\#) it receives from the SRT. 84\% of MISTRAL detectors are alive and usable. The average optical efficiency of the receiver was measured to be $\simeq35\%$. The figure of merit for the sensitivity of the KIDs is their Noise Equivalent Power (NEP) which represents the incoming power which produces a signal equal to the noise power spectrum of the KIDs. In Fig. \ref{fig:cal} (right panel) we show an histogram of the resulting measurement which shows a median value of $8.07 \times10^{-16}W/\sqrt{Hz}$. 

MISTRAL receiver was transported and installed at the focus of the SRT between May and June 2023 (see Fig. \ref{fig:cal}, left panel).
The aforementioned NEP's, nominally would translate into a $NEFD \simeq 2.8mJy  \sqrt{s}$ \cite{perotto20}. However, what is not taken into account in this estimate is the telescope efficiency and the noise added by the atmospheric fluctuation. We thus have undertaken a realistic simulation which assumes an arbitrary telescope efficiency of 30\%, and takes into account the real atmospheric noise at the SRT observatory at 22GHz, and then extrapolated it to 90GHz using $am$ code\footnote{https://lweb.cfa.harvard.edu/~spaine/am/}. This results into an approximate $NEFD \simeq 10-15mJy  \sqrt{s}$. Assuming the definition reported by Perotto et al. 2020 \cite{perotto20}, we extracted a mapping speed of $MS= 380'^{2}/mJy^{2}/h$ \cite{isopi23}.

\section{Conclusions}
\label{concl}

The full comprehension of the matter distribution around the universe is crucial both for cosmology and for astrophysics. The Sunyaev Zel'dovich effect is a powerful tool to study low density environments and search for bridges and filaments in the cosmic web. High angular resolution is crucial to understand and map galaxy clusters and the surrounding medium. We developed MISTRAL, which, coupled with the SRT, is an ideal instrument to map the sky, at 90GHz, with $12''$ angular resolution. MISTRAL is a cryogenic camera with an array of KIDs cooled down at $\simeq 200mK$. We recently installed the camera at the Gregorian room of the SRT and soon we expect to open it to the sky for the first light.


%
%
%

\end{document}